\documentclass[aps,amsfonts,showpacs,twocolumn,floatfix]{revtex4}
\usepackage{epsfig,amsopn,amssymb,amsmath,amsfonts}
\usepackage{graphicx}
\begin{document}
\def\be{\begin{equation}}
\def\ee{\end{equation}}
\def\bea{\begin{eqnarray}}
\def\eea{\end{eqnarray}}
\def\bef{\begin{figure}}
\def\eef{\end{figure}}
\def\l{\label}
\def\fr{\frac}
\def\th{\theta}
\def\o{\omega}
\def\O{\Omega}
\def\eps{\epsilon}
\def\p{\partial}
\title{Slow relaxation in long-range interacting systems with stochastic dynamics}
\author{Shamik Gupta and David Mukamel}
\affiliation{
Physics of Complex Systems, Weizmann Institute of Science, Rehovot 76100, Israel}
\date{\today}
\begin{abstract}
Quasistationary states are long-lived nonequilibrium states, observed in some systems with long-range interactions under deterministic Hamiltonian evolution. These intriguing non-Boltzmann states relax to equilibrium over times which diverge algebraically with the system size. To test the robustness of this phenomenon to non-deterministic dynamical processes, we have generalized the paradigmatic model exhibiting such a behavior, the Hamiltonian Mean-Field model, to include energy-conserving stochastic processes. Analysis, based on the Boltzmann equation, a scaling approach and numerical studies, demonstrates that in the long time limit, the system relaxes to the equilibrium state on timescales which do not diverge algebraically with the system size. Thus, quasistationarity takes place only as a crossover phenomenon on times determined by the strength of the stochastic process.    
\end{abstract}
\maketitle
Systems with long-range interactions have been a subject of extensive studies in recent years. In these systems, the inter-particle potential at large separation, $r$, decays slower than $1/r^d$ in $d$ dimensions; for reviews, see \cite{review}. Examples are self-gravitating systems \cite{Paddy:1990}, plasmas \cite{Nicholson:1992}, dipolar magnets \cite{Landau:1960} and wave-particle interacting systems \cite{Barre:2004}. Long-range interactions lead to non-additivity, whereby thermodynamic quantities scale superlinearly with the system size. This often results in unusual features such as a negative microcanonical specific heat \cite{Lynden-Bell-Thirring:196870}, inequivalence of statistical ensembles \cite{Barre:2001, Mukamel:2005}, and many others \cite{Bouchet:2005}. 

Models with long-range interactions often exhibit striking dynamical features like slow relaxation \cite{Mukamel:2005, Chavanis:2002} and broken ergodicity \cite{Borgonovi:2004, Mukamel:2005, Bouchet:2008}. A very interesting characteristic feature is the occurrence of long-lived nonequilibrium quasistationary states (QSS). These states involve a slow relaxation of macroscopic observables over times which diverge algebraically with the system size. This suggests that in the thermodynamic limit, the system never reaches the Boltzmann equilibrium, instead remains trapped in the QSS. These intriguing states are usually observed under deterministic Hamiltonian evolution. In reality, stochastic dynamical moves resulting from coupling either to external environment or to internal degrees of freedom are often present. It is thus of interest to analyze the robustness of QSS to stochasticity. The aim of this work is to explore their existence under \textit{stochastic} dynamics beyond deterministic evolution.  

A prototypical model which allows for detailed studies of QSS is the Hamiltonian Mean-Field (HMF) model. The model describes $N$ globally coupled $XY$ spins with the Hamiltonian \cite{Ruffo:1995}:
\be
H=\sum_{i=1}^{N}\fr{p_i^2}{2}+\fr{1}{2N}\sum_{i,j=1}^{N}\left[1-\cos(\th_i-\th_j)\right],
\l{HMF-H}
\ee
where $\th_i \in [0,2\pi)$ is the phase of the $i$-th spin and $p_i$ its conjugate momentum. The model represents physical systems like gravitational sheet models \cite{Tsuchiya:1994} and the free-electron laser \cite{Barre:2004}. Within a microcanonical ensemble, the time evolution of the system is described by the deterministic Hamilton equations: 
\be
\fr{d\th_i}{dt}=p_i, ~~~~\fr{dp_i}{dt}=-m_x\sin \th_i+m_y\cos \th_i. 
\l{hameq}
\ee
Here, $m_x$ and $m_y$, respectively, are the $x$ and the $y$ components of the specific magnetization vector, $\vec{m}=\sum_{i=1}^{N}(\cos \th_i,\sin \th_i)/N$. The dynamics, Eq. (\ref{hameq}), conserves energy and momentum. Defining the temperature $T$ as twice the specific kinetic energy, the energy per particle, $\eps$, satisfies the relation $\eps=T/2+(1-m^2)/2$, where $m^2=m_x^2+m_y^2$ and the Boltzmann constant is taken to be unity. In equilibrium, the model shows a continuous transition from a paramagnetic to a ferromagnetic phase at the critical energy $\eps_c=3/4$, corresponding to the critical temperature $T_c=1/2$ \cite{Ruffo:1995}, \cite{Barre:2005}.

In the HMF model, a typical initial state to study relaxation to equilibrium is one which is homogeneous in angles and uniform in momenta in some momentum interval (the ``water-bag'' initial condition). It is observed that, at an energy interval just below $\eps_c$ and under the dynamics of Eq. (\ref{hameq}), the magnetization stays close to its initial value over a long time which scales with the system size as $ N^\delta$, where $\delta >1$ \cite{Yamaguchi:2004, Bouchet:2005rapid}. Recent studies of QSS have revealed many interesting features, e.g., anomalous diffusion \cite{Latora:1999}, non-Gaussian velocity distributions \cite{Latora:2001}, and vanishing Lyapunov exponents \cite{Latora:2001}.

Recent studies have invoked a coupling of the HMF system to an external heat bath, thereby allowing for energy exchange between the two \cite{Baldovin:20069}. These studies suggest that QSS may occur, depending on the interplay of different timescales underlying the coupling to the bath. In particular, it has been shown that the coupling to the bath results in relaxation to the canonical Gibbs-Boltzmann equilibrium state on a timescale which does not diverge with the system size. Since quasistationarity has, so far, been observed only in isolated systems with deterministic, energy-conserving dynamics, testing its stability to stochastic energy-conserving processes, where no external bath is involved, would be of great interest.   

In this Letter, we address the question of robustness of QSS with respect to \textit{stochastic} dynamics of an isolated system within a \textit{microcanonical ensemble}, where the energy is conserved. To this end, we generalize the HMF model to include stochastic dynamical moves in addition to the deterministic ones, Eq. (\ref{hameq}). We study our model by analyzing the Boltzmann equation for the time evolution of the phase space density, and also by a scaling approach and by extensive numerical simulations. We provide physical arguments to suggest that the stochastic process introduces a cut-off in the relaxation time of the QSS. As a result, QSS are maintained only as a crossover phenomenon over a characteristic time, which is determined by the strength of the stochastic process. In particular, at long times, the relaxation time \textit{does not} any more scale algebraically with the system size. Consequently, there are \textit{no} quasistationary states. Our scaling form for the relaxation time is in very good agreement with results from our numerical simulations.

We now define our generalized HMF model. It follows a piecewise deterministic dynamics, whereby the Hamiltonian evolution, Eq. (\ref{hameq}), is randomly interrupted by stochastic inter-particle collisions that conserve energy and momentum. We consider collisions in which only the momenta are updated stochastically. Since the momentum variable in the HMF model is one-dimensional, and there are two conservation laws for momentum and energy, one has to resort to three-particle collisions. Namely, three random particles, $(i,j,k)$, collide and their momenta are updated stochastically, $(p_i,p_j,p_k) \rightarrow (q_i,q_j,q_k)$, while conserving energy and momentum and keeping the phases unchanged. Thus, the model evolves under the following repetitive sequence of events: deterministic evolution, Eq. (\ref{hameq}), for a time interval whose length is exponentially distributed, followed by a single sweep of the system for three-particle collisions, which consists of $N^3$ collision attempts. 

We proceed by considering the Boltzmann equation of our model. In the limit $N \rightarrow \infty$, this equation governs the time evolution of the single-particle phase space distribution $f(\th,p,t)$, and is given by:
\bea
&&\fr{\p f}{\p t}+p\fr{\p f}{\p \th}-\fr{\p \langle v \rangle}{\p \th}\fr{\p f}{\p p}=\left(\fr{\p f}{\p t}\right)_c,\l{Boltz}\\
&&\left(\fr{\p f}{\p t}\right)_c=\int d\eta R[f(\th,q,t)f(\th',q',t)f(\th'',q'',t)\nonumber \\
&&~~~~~~~~~~~~~-f(\th,p,t)f(\th',p',t)f(\th'',p'',t)], \l{collisionterm} \\
&&R=\alpha\delta(p+p'+p''-q-q'-q'')\nonumber \\
&&~~~~~~\delta\left(\fr{1}{2}(p^2+p'^2+p''^2)-\fr{1}{2}(q^2+q'^2+q''^2)\right),
\eea
where $d\eta \equiv dp'dp''dqdq'dq''d\th'd\th''$. In Eq. (\ref{Boltz}), $\langle v\rangle=\int dp'\int d\th'\left[1-\cos(\th-\th')\right]f(\th',p',t)$ is the average potential. Equation (\ref{collisionterm}) represents the three-body collision term, and $R$ is the rate for collisions $(p,p',p'') \rightarrow (q,q',q'')$ that conserve energy and momentum. The constant $\alpha$ has the dimension of 1/time and sets the scale for collisions: on average, there is one collision after every time interval $\alpha^{-1}$. The Boltzmann equation with similar three-particle collisions in one dimension was considered earlier \cite{Ma:1983}. We refer to the Boltzmann equation with $\alpha=0$ as the Vlasov-equation limit \cite{Nicholson:1992}. Note that both the Boltzmann and the Vlasov equations are valid for infinite $N$ and have size-dependent correction terms when $N$ is finite.

Significant physical insight into the existence of QSS can be gained by a direct inspection of Eq. (\ref{Boltz}). In the Vlasov limit, corresponding to the deterministic dynamics, any state which is homogeneous in angles but with arbitrary momentum distribution is stationary. In this limit, it has been shown that QSS are related to the linear stability of the stationary solutions, chosen as the initial state \cite{Yamaguchi:2004, Jain:2007}. For example, consider an initial state which is homogeneous in angles and uniform in momenta over $[-p_0,p_0]$, where $p_0=\sqrt{6\eps-3}$. Such a state was shown to be linearly stable for energies in the range $\eps^* \equiv 7/12 < \eps < \eps_c$, and unstable for $ \eps < \eps^*$ \cite{Yamaguchi:2004, Jain:2007}. As a result, QSS are observed when the energy lies in the range $\eps^* < \eps < \eps_c$. In a finite system, such states finally relax to equilibrium due to finite-size effects which come into play over a time $\sim N^\delta$, where $\delta >1$ \cite{Bouchet:2005rapid}. For example, for $\eps=0.69$, numerics gives $\delta \simeq 1.7$ \cite{Yamaguchi:2004}. AT long-times, one has
\be
m(t) \sim \fr{1}{\sqrt{N}}e^{t/N^\delta}; ~~~~~~~~ t \gg N^\delta,
\l{stable}
\ee 
where the prefactor accounts for fluctuations in the initial state. For $\eps < \eps^*$, linear instability results in a faster relaxation towards equilibrium as
\be
m(t) \sim \fr{1}{\sqrt{N}}e^{\gamma t}; ~~~~~~~~ t \gg \fr{1}{\gamma},
\l{unstable}
\ee
where $\gamma^2=6\left(\fr{7}{12}-\eps\right)$ is independent of $N$ \cite{Jain:2007}. Thus, there are no QSS for energies below $\eps^*$. On the other hand, a homogeneous state with Gaussian-distributed momenta is linearly unstable at any energy below $\eps_c$, and no QSS are observed \cite{Jain:2007}.

Let us now turn to a discussion of QSS in the generalized HMF model, i.e., under noisy microcanonical evolution, in the light of the Boltzmann equation. First, we note that, unlike the Vlasov equation, a homogeneous state with an arbitrary momentum distribution is not stationary under the Boltzmann equation; instead, only a Gaussian distribution is stationary. Suppose we start with an initial homogeneous state with uniformly distributed momenta. Then, under the dynamics, the momentum distribution will evolve towards the stationary Gaussian distribution. Interestingly, although the momentum distribution evolves, the initial $\th$-distribution does not change in time. This is a result of the product-measure form of the initial distribution.

In a finite system, however, there are fluctuations in the initial state. These fluctuations make the homogeneous state with Gaussian-distributed momenta linearly unstable under the Boltzmann equation at all energies $< \eps_c$, as we demonstrate below. This results in a fast relaxation towards equilibrium. 

The proof of linear instability of a homogeneous state with Gaussian-distributed momenta at energies below $\eps_c$ was performed in the Vlasov limit in Ref. \cite{Yamaguchi:2004, Jain:2007}. We have generalized this analysis to the case of the Boltzmann equation \cite{preparation}. The essential steps of the analysis are summarized below, for the simple case of energies just below the critical point.

The stability analysis is carried out by linearizing Eq. (\ref{Boltz}) about the homogeneous state. We expand $f(\th,p,t)$ as $f(\th,p,t)=f^{(0)}(p)[1+\lambda f^{(1)}(\th,p,t)]$ with $f^{(0)}(p)=e^{-p^2/2T}/(2\pi\sqrt{2\pi T})$. Here, since the initial angles and momenta are sampled independently according to $f^{(0)}(p)$, fluctuations for finite $N$ make the small parameter $\lambda$ of $O(1/\sqrt{N})$. Next, one assumes that, at long times, the dynamics is dominated by the eigenmode with the largest eigenvalue of the linearized Boltzmann equation, so that $f^{(1)}(\th,p,t)=f^{(1)}_{k}(p,\o)e^{i(k\th+\o t)}$. Since the average potential $\langle v \rangle$ in Eq. (\ref{Boltz}) involves $e^{\pm i\th}$, one needs to consider only $k=\pm 1$. The coefficients $f^{(1)}_{\pm 1}$ then satisfy \cite{preparation}
\bea
&&\!\!\!\!\!\!\!\!\pm ipf^{(1)}_{1\pm}(p,\o)\mp \fr{2\pi}{2if^{(0)}}\fr{\p f^{(0)}}{\p p}\int dp'f^{(0)}(p')f^{(1)}_{\pm 1}(p',\o)\nonumber \\
&&\!\!\!\!\!\!\!\!+\alpha(4\pi)^2\int dp'dp''dqdq'dq''Rf^{(0)}(p')f^{(0)}(p'')\nonumber \\
&&\!\!\!\!\!\!\!\!\times [f^{(1)}_{\pm 1}(p,\o)-f^{(1)}_{\pm 1}(q,\o)]=-i\o f_{1\pm}(p,\o).
\l{lin2}
\eea
Treating $\alpha$ as a small parameter, we solve the above equation perturbatively in $\alpha$. In the absence of collisions ($\alpha=0$), the above analysis reduces to that of the Vlasov equation and to the unperturbed solutions, namely, the frequencies $\o^{(0)}$ and the coefficients $f^{(1)}_{\pm 1}(q,\o^{(0)})$, which are obtained from the analysis in Ref. \cite{Jain:2007}. In particular, slightly below the critical point $\eps_c$, the unperturbed real frequencies $\O^{(0)}=i\o^{(0)}$ are given by $|\O^{(0)}| \approx \fr{2}{\sqrt{\pi}}(T_c-T)$. Thus, in the Vlasov limit, the homogeneous state with Gaussian-distributed momenta is unstable below the critical energy. To obtain the perturbed frequencies $\O$ to lowest order in $\alpha$, we now substitute the unperturbed solutions into Eq. (\ref{lin2}). After a straightforward, but lengthy algebra, one obtains, at an energy slightly below the critical point, the perturbed frequencies to be given by \cite{preparation}
\be
\O \approx |\O^{(0)}|[1+\alpha A], \mathrm{~with~} A=\fr{2\pi^{3/2}}{\sqrt{3}}\left(1-\fr{1}{\sqrt{5}}\right).
\ee
This equation suggests that, to leading order in $\alpha$, the frequencies $\O$ are real for energies just below the critical value, and vanish at the critical point. Thus, a homogeneous state with Gaussian-distributed momenta is linearly unstable under the Boltzmann equation at energies just below the critical point, and neutrally stable at the critical point.

Armed with the above background, we can now analyze the evolution of magnetization in our model while starting from a water-bag initial condition. The two time scales which govern the evolution of the  magnetization are (i) the scale over which collisions occur, given by $\alpha^{-1}$, and (ii) the scale $\sim N^\delta$, over which finite-size effects add corrections to the Boltzmann equation. The interplay between the two timescales may be naturally analyzed by invoking a scaling approach, as we demonstrate below. 

\bef
\includegraphics[scale=0.4]{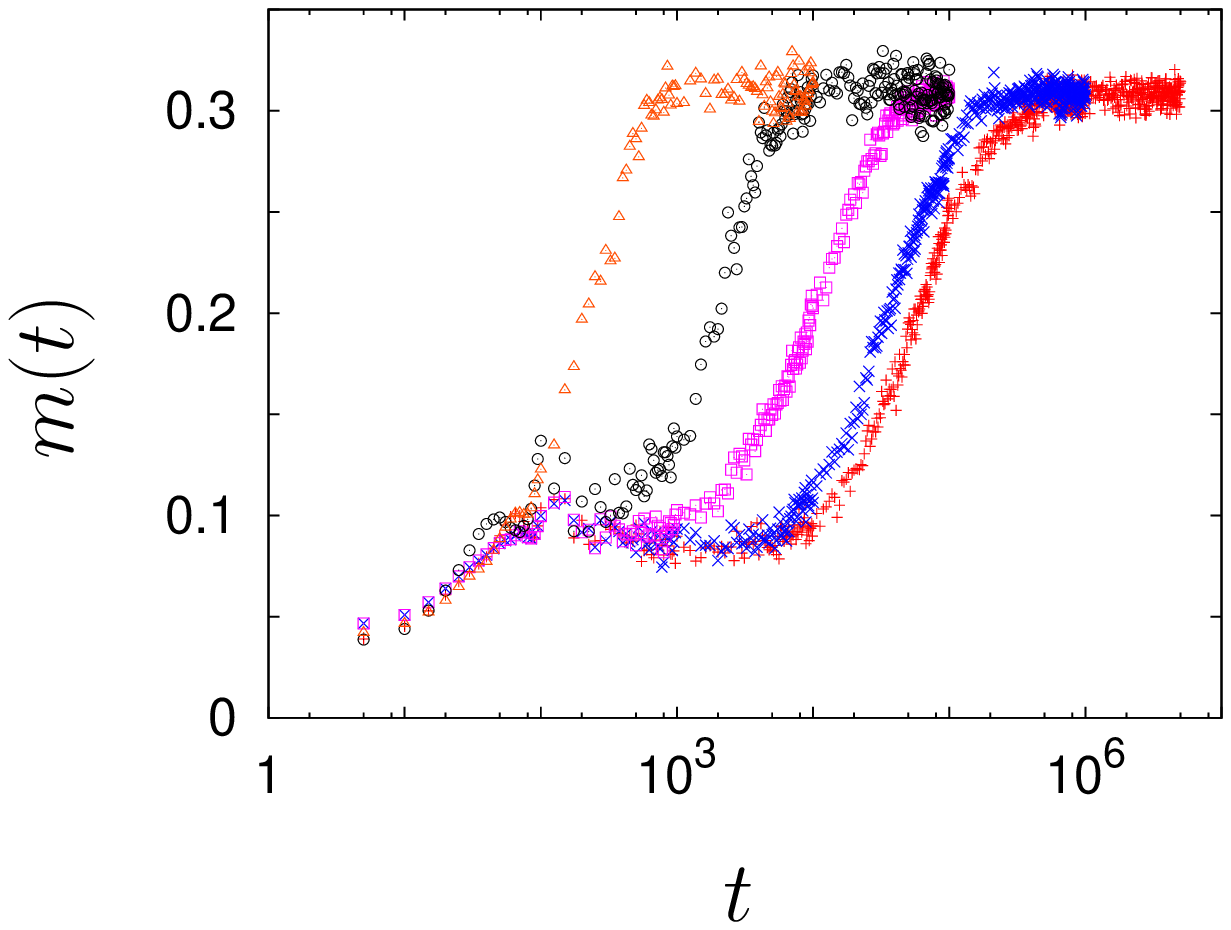}
\caption{(Color online) Magnetization vs. time for $N=500$ at $\eps=0.69$ and for $\alpha$ values (right to left) $10^{-5}, 10^{-4}, 10^{-3},10^{-2},10^{-1}$. Data averaging has been typically over hundred histories. With increasing $\alpha$, one can observe a faster relaxation towards equilibrium.}
\l{fig1}
\eef

For $\alpha^{-1} \ll N^\delta$, and times $\alpha^{-1} \ll t \ll N^\delta$, the system size is effectively infinite and the evolution follows the Boltzmann equation. Here, frequent collisions at short times drive the momentum distribution towards a Gaussian. As noted above, until this happens, the initial magnetization does not change in time. Over the time the momentum distribution becomes Gaussian, the instability of such a state under the Boltzmann equation leads to a fast relaxation towards equilibrium, similar to the result for the Vlasov-unstable regime given in Eq. (\ref{unstable}). The asymptotic behavior of the magnetization is thus
\be
m(t) \sim \fr{1}{\sqrt{N}}e^{\alpha t}; ~~~~~~~~ N^\delta \gg t \gg \alpha^{-1}.
\l{mag1}
\ee
Requiring that $m(t)$ acquires a value of $O(1)$, the above equation gives the relaxation time $\tau_\mathrm{S}$, determined by the stochastic process, as $\tau_\mathrm{S} \sim \ln N/\alpha$.

\bef
\includegraphics[scale=0.4]{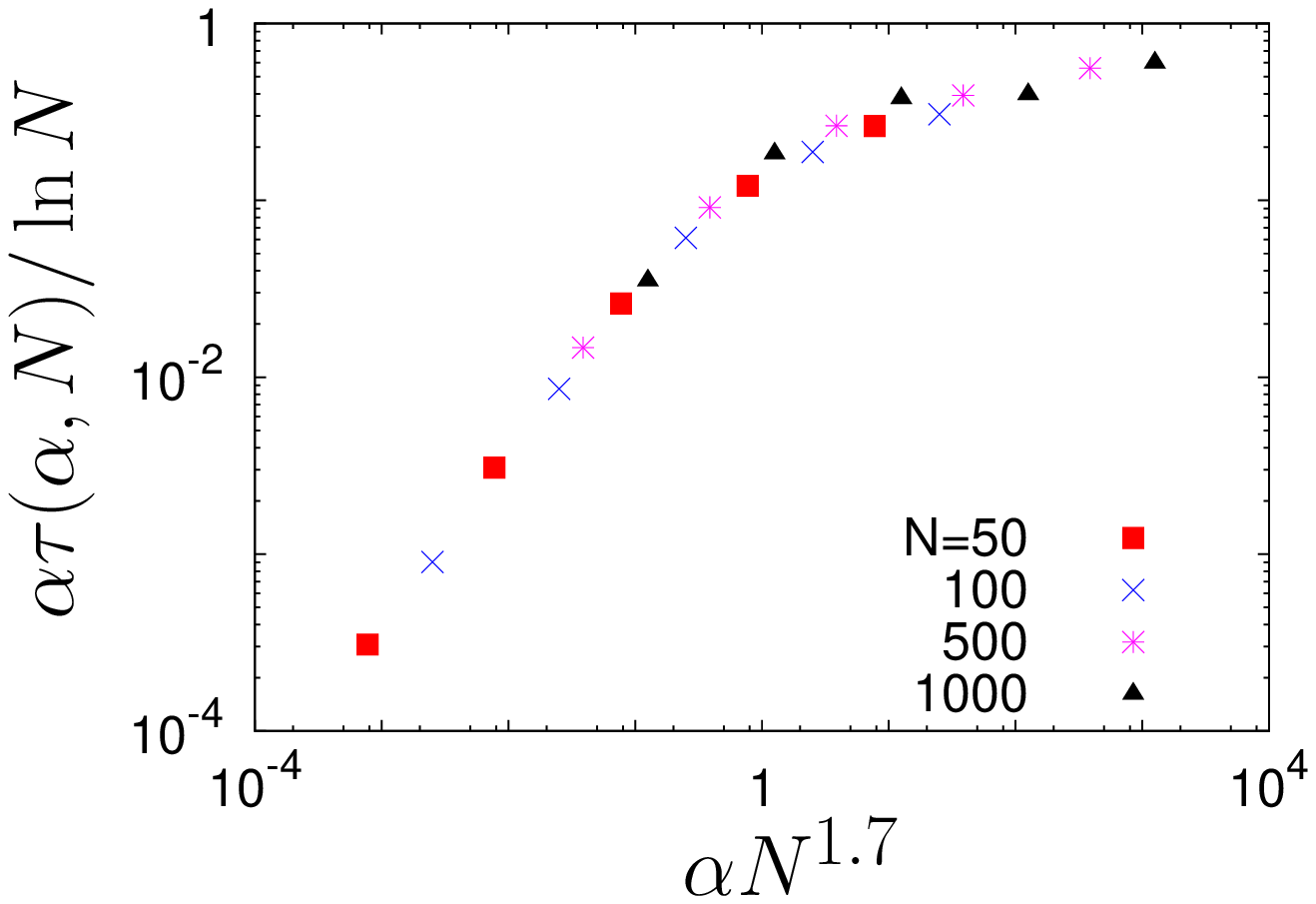}
\caption{(Color online) $\alpha \tau(\alpha, N)/\ln N$ vs. $\alpha N^\delta$, showing scaling collapse in accordance with Eq. (\ref{scaling}). Here, $\eps=0.69$. Data averaging varies between $5 \times 10^4$ histories for the smallest system and $100$ histories for the largest one.}
\l{fig2}
\eef

In the opposite limit, $\alpha^{-1} \gg N^\delta$, collisions are infrequent, and therefore, the process that drives the momentum distribution to a Gaussian is delayed. The magnetization stays close to its initial value, and relaxes only over the time $\sim N^\delta$, over which finite-size effects come into play. Here, similar to the result for the Vlasov-stable regime in Eq. (\ref{stable}), the magnetization at late times behaves as
\be
m(t) \sim \fr{1}{\sqrt{N}}e^{t/N^\delta}; ~~~~~~~~\alpha^{-1} \gg t \gg N^\delta.
\l{mag2}
\ee
This equation gives the relaxation time $\tau_\mathrm{D}$, determined by the deterministic process, as $\tau_\mathrm{D} \sim N^\delta \ln N$.

Interpolating between the above two limits of the timescales, one expects the relaxation time $\tau(\alpha,N)$ to obey, $\tau^{-1}=\tau_\mathrm{S}^{-1}+\tau_\mathrm{D}^{-1}$, yielding 
\be
\tau(\alpha, N) \sim \fr{\ln N}{\alpha + 1/N^\delta}.
\l{T}
\ee
More generally, Eq. (\ref{T}) suggests a scaling form
\be
\tau(\alpha, N) \sim \fr{\ln N}{\alpha}g(\alpha N^{\delta}),
\l{scaling}
\ee
where, consistent with Eqs. (\ref{mag1}) and (\ref{mag2}), the scaling function $g(x)$ behaves as follows: $g(x) \sim x$ for $x \ll 1$, while $g(x) \rightarrow$ constant for $x \gg 1$.

Equation (\ref{scaling}) implies that, for fixed $N$, the relaxation time of the water-bag initial state exhibits a crossover, from being of order $N^\delta \ln N$ (corresponding to QSS) for $\alpha \ll 1/N^\delta$, to being of order $\ln N$ for $\alpha \gg 1/N^\delta$. This brings us to the main conclusion of this work: In the presence of collisions, the relaxation at long times does not occur over an algebraically growing timescale. This implies that, under noisy microcanonical evolution, QSS occur only as a crossover phenomenon, and are lost in the limit of long times. 

To verify the above predictions, in particular, the scaling form in Eq. (\ref{scaling}), we performed extensive numerical simulations of our model. The Hamilton equations, Eq. (\ref{hameq}), were integrated using a symplectic fourth-order integrator. In realizing the stochastic process $(p,p',p'') \rightarrow (q,q',q'')$ while conserving the three-particle energy $E$ and momentum $P$, we note that the updated momenta lie on a circle formed by the intersection of the plane $p+p'+p''=P$ and the spherical surface $p^2+p'^2+p''^2=2E$. The radius of this circle is given by $r=\sqrt{2E-P^2/3}$. The new momenta may thus be parametrized in terms of an angle $\phi$ measured along this circle, as $q=\fr{P}{\sqrt{3}}+r\sqrt{\fr{2}{3}}\cos \phi, q'=\fr{P}{\sqrt{3}}-\fr{r}{\sqrt{6}}\cos \phi-\fr{r}{\sqrt{2}}\sin \phi, q''=\fr{P}{\sqrt{3}}-\fr{r}{\sqrt{6}}\cos \phi+\fr{r}{\sqrt{2}}\sin \phi$ \cite{Ma:1983}. Stochasticity in updates is achieved through choosing the angle $\phi$ uniformly in $[0,2\pi)$.

Following the above scheme, typical time evolutions of the magnetization in our model for $N=500$ and several values of $\alpha$ at an energy density $\eps=0.69$ are shown in Fig. \ref{fig1}. The relaxation time $\tau(\alpha,N)$ is taken as the time for the magnetization to reach the fraction $0.8$ of the final equilibrium value (the result, however, is not sensitive to this choice). At $\eps=0.69$, where the equilibrium value of the magnetization is $\simeq 0.3$ and $\delta \simeq 1.7$ \cite{Yamaguchi:2004}, we plot $\alpha \tau(\alpha, N)/\ln N$ vs. $\alpha N^\delta$ to check the scaling form in Eq. (\ref{scaling}). Figure \ref{fig2} shows an excellent scaling collapse over several decades. This is consistent with our prediction for QSS as a crossover phenomenon under noisy microcanonical dynamics.

In summary, a generalized HMF model, which evolves with a piecewise stochastic microcanonical dynamics, is introduced and analyzed for the existence of quasistationary states. It is shown that the quasistationarity existing with pure deterministic dynamics is lost in the presence of stochastic dynamics in the long time limit. It occurs only as a crossover phenomenon on times determined by the strength of the stochastic process. It would be of great interest to explore the general validity of this result in other models with long-range interactions.  

We thank O. Cohen, T. Dauxois, O. Hirschberg, H. Posch and S. Ruffo for fruitful discussions and comments. The support of the Israel Science Foundation (ISF) and the Minerva Foundation with funding from the Federal German Ministry for Education and Research is gratefully acknowledged.


\begin{thebibliography}{9}
\bibitem{review}A. Campa, T. Dauxois and S. Ruffo, Phys. Rep. {\bf 480}, 57 (2009); F. Bouchet, S. Gupta and D. Mukamel, Physica A (in press); also eprint:arXiv:cond-mat/1001.1479.
\bibitem{Paddy:1990}T. Padmanabhan, Phys. Rep. {\bf 188}, 285 (1990).
\bibitem{Nicholson:1992}D. R. Nicholson, {\em Introduction to Plasma Physics} (Krieger Publishing Company, Florida, 1992).
\bibitem{Landau:1960}L. D. Landau and E. M. Lifshitz, {\em Electrodynamics of Continuous Media} (Pergamon, London, 1960).
\bibitem{Barre:2004}J. Barr\'{e} \textit{et al.}, Phys. Rev. E {\bf 69}, 045501(R) (2004).
\bibitem{Lynden-Bell-Thirring:196870}D. Lynden-Bell and R. Wood, Mon. Not. R. Astron. Soc. {\bf 138}, 495 (1968); W. Thirring, Z. Phys. {\bf 235}, 339 (1970).
\bibitem{Barre:2001}J. Barr\'{e}, D. Mukamel and S. Ruffo, Phys. Rev. Lett. {\bf 87}, 030601 (2001).
\bibitem{Mukamel:2005}D. Mukamel, S. Ruffo and N. Schreiber, Phys. Rev. Lett. {\bf 95}, 240604 (2005).
\bibitem{Bouchet:2005}F. Bouchet and J. Barr\'{e}, J. Stat. Phys. {\bf 118}, 1073 (2005).
\bibitem{Chavanis:2002}P. H. Chavanis, in {\it Dynamics and Thermodynamics of Systems with Long-Range Interactions}, edited by T. Dauxois, S. Ruffo, E. Arimondo and M. Wilkens (Springer-Verlag, Berlin, 2002), Vol. 602. 
\bibitem{Borgonovi:2004}F. Borgonovi \textit{et al.}, J. Stat. Phys. {\bf 116}, 1435 (2004).
\bibitem{Bouchet:2008}F. Bouchet \textit{et al.}, Phys. Rev. E, {\bf 77}, 011125 (2008).
\bibitem{Ruffo:1995}M. Antoni and S. Ruffo, Phys. Rev. E {\bf 52}, 2361 (1995).
\bibitem{Tsuchiya:1994}T. Tsuchiya, T. Konishi and N. Gouda, Phys. Rev. E {\bf 50}, 2607 (1994).
\bibitem{Barre:2005}J. Barr\'{e} \textit{et al.}, J. Stat. Phys. {\bf 119}, 677 (2005).
\bibitem{Yamaguchi:2004}Y. Y. Yamaguchi \textit{et al.}, Physica A {\bf 337}, 36 (2004).
\bibitem{Bouchet:2005rapid}F. Bouchet and T. Dauxois, Phys. Rev. E {\bf 72}, 045103(R) (2005).
\bibitem{Latora:1999}V. Latora, A. Rapisarda and S. Ruffo, Phys. Rev. Lett. {\bf 83}, 2104 (1999).
\bibitem{Latora:2001}V. Latora, A. Rapisarda and C. Tsallis, Phys. Rev. E {\bf 64}, 056134 (2001).
\bibitem{Baldovin:20069}F. Baldovin and E. Orlandini, Phys. Rev. Lett. {\bf 96}, 240602 (2006); {\bf 97}, 100601 (2006); F. Baldovin, P. H. Chavanis and E. Orlandini, Phys. Rev. E {\bf 79}, 011102 (2009).
\bibitem{Ma:1983}S-K. Ma, J. Stat. Phys. {\bf 31}, 107 (1983).
\bibitem{Jain:2007}K. Jain, F. Bouchet and D. Mukamel, J. Stat. Mech.: Theory Exp. P11008 (2007).
\bibitem{preparation}S. Gupta and D. Mukamel (To be published).
\end{thebibliography}
\end{document}